\documentclass[preprint,aps,showpacs]{revtex4}

\usepackage{amsmath}
\usepackage{amssymb}
\usepackage{bbold}
\usepackage{mathrsfs}
\usepackage{graphicx, psfrag}
\usepackage[centerlast]{caption}
\usepackage{float}
\usepackage{subfig}
\usepackage{tikz}
\usepackage{pgfplots}
\usepackage[colorlinks=true, citecolor=blue, urlcolor = blue, linkcolor= red, bookmarks=true]{hyperref}
\usepackage{epstopdf}
\usepackage{multirow}

\begin{document}
%%--------------------------- Shortcuts for various Macros -------------------- %%
\def \beq{\begin{equation}}
\def \eeq{\end{equation}}
\def \bse{\begin{subequations}}
\def \ese{\end{subequations}}
\def \bea{\begin{eqnarray}}
\def \eea{\end{eqnarray}}
\def \bem{\begin{displaymath}}
\def \eem{\end{displaymath}}
\def \bem{\begin{pmatrix}}
\def \eem{\end{pmatrix}}
\def \bb{\bibitem}
\def \bs{\boldsymbol}
\def \nn{\nonumber}
\def \mf{\tilde{J}_1}
\def \mj{\tilde{J}_0}
\def \mh{\mathcal{H}}
\def \ma{\mathcal{A}}
\def \md{\mathcal{D}}
\def \mg{\mathcal{G}}
\def \ha{\hat{a}}
\def \hs{\hat{S}}
\def \hh{\hat{H}}
\def \mH{\hat{\mathcal{H}}}
\def \hp{\hat{\Psi}}
\def \hb{\hat{b}}
\def \hB{\hat{\mathcal{T}}}
\def \hN{\hat{\mathcal{N}}}
%%------------------- New Commands for various operators -----------------%%
\newcommand{\cc}{\color{red}}
\newcommand{\cb}{\color{blue}}
\newcommand{\upa}{\uparrow}
\newcommand{\dna}{\downarrow}
\newcommand{\pdag}{\phantom\dagger}
\newcommand{\bra}[1]{\langle #1 |}
\newcommand{\ket}[1]{| #1 \rangle}
\newcommand{\braket}[2]{\langle #1 | #2 \rangle}
\newcommand{\ketbra}[2]{| #1 \rangle \langle #2 |}
\newcommand{\expect}[1]{\langle #1 \rangle}
\newcommand{\inx}[1]{\int d\bs x \Big [ #1 \Big ]}
\newcommand{\si}[1]{\hp^{\pdag}_{#1} (\bs x)}
\newcommand{\dsi}[1]{\hp^\dag_{#1} (\bs x)}
\newcommand{\evolve}[1]{\frac{\partial #1}{\partial t}}
%---------------------------------------------------------------%
\title{\textbf{Synthetic Gauge Fields for Ultra Cold atoms: A Primer}}
 \author{ Sankalpa Ghosh}
\affiliation{Department of Physics,  Indian Institute of Technology Delhi, Hauz Khas, New Delhi-110016, India}
\author{Rashi Sachdeva}
\affiliation{Department of Physics, Indian Institute of Technology Kanpur, Kanpur-208016, India}
\email{sankalpa@physics.iitd.ac.in}
\begin{abstract}
We start by reviewing the concept of gauge invariance in quantum mechanics, for Abelian and Non-Ableian cases.
Then we idescribe how the various gauge potential and field can be associated with the geometrical 
phase acquired by a quantum mechanical wave function while adiabatically evolving in a parameter space. Subsequently we show how this concept is exploited to generate light induced gauge field for neutral ultra cold bosonic atoms. As an example of such light induced Abelian and Non Abelian gauge field for ultra cold atoms we disucss ultra cold atoms in a rotating trap and creation of synthetic spin orbit coupling for ultra cold atomic systems using Raman lasers.
\end{abstract}
\maketitle

\tableofcontents
%%%%%%%%%%%%%%%%%%%%%%%%%%%%%%%%%%%
\section{Introduction}
"Quantum Simulation with ultra cold atoms" \cite{QS} is a much discussed and investigated topic nowadays. Such quantum simulation implies that certain model hamiltonians that were originally proposed 
to explain exotic quantum behavior in condesned matter systems and sometimes in high energy physics can be  realized with ultra cold atoms in various type of optical potentials. 
In a typical condensed matter or high energy systems, the actual system is far more complex than one described by such model hamiltonians and that leaves a lot of issues less than verified. Ultra cold atomic systems on the otherhand much more controlled and model behavior can be studied almost exactly. 
In nature Electromagnetic fields are known as Abelian gauge field whereas more complicated Non Abelian gauge fields are the one that are responsible for strong and weak interaction. Gauge fields are therefore known and confirmed to define three fundamental interaction formed the very basis of our understanding of the
microscopic world \cite{FTB}. Given this crucial role of gauge fields in fundamental physics
the recent success in simulation of Abelian and Non Abelian gauge fields for ultra cold neutral atoms 
\cite{Dalibard, Ketterle, Cornell, lin1, lin2}
 is one of the most significant achievment in the field of ultra cold atomic research. 

Apart from allowing us studying the fields responsible for fundamental interaction in a non-trivial context this development  also opens the possibility of studying the model hamiltonians that predicts 
Quantum Hall Effect \cite{QH}, Topological Insulators and Superconductors \cite{TI}, Interesting vortex phases in superconductor \cite{VL} and other interesting magnetic field dependent phenomena in electronic system. There have been already a number of excellent and detailed review articles written by some of the most prominent experts in this field. We 
are not going to repeat the topic of their review. What we want to describe the basic physics associated with the processes that creates such synthetic abelian and non abelian gauge field for ultra cold atomic system and a systematic comparison with true gauge fields 
that occurs in nature and responsible for the fundamental interactions with the aim to  analyze  their "fundamental"-ness. 

To that prupose we shall  first begin with the Gauge invarinace of Schr\"odinger Equation ( for example see \cite{JJS}) for Abelian Gauge theory in Sec. \ref{Abl}. In the next section \ref{NonAb} we shall introduce the Non Abelian gauge theory. Rather than taking a field theory route we shall do that in a theoretical background of Quantum Mechanics \cite{Quantum}. In the next section \ref{sec:Berry} we shall show how such gauge fields described in the two preceeding section can arise in ordinary Quantum Mechanics purely for geometrical reasons.
In the next section \ref{RotBose} we shall show how in rotatating Bose Einstein Condensate such an abelian gauge field can be simulated. The section \ref{ColdBerry} describes a general scheme of producing Abelian and Non Abelian gauge potential for multilevel ultra cold atoms by using laser induced coupling between different hyperfine states. In the next section we shall try to describe how both the "fundamental gauge fields" as well as their synthetic counterpart in ultra cold atomic systems known as geometric gauge potentials to a more fundamental geometrical concept of parallel transport. We shall also briefly describe in this section why inspite of 
this connection the gauge fieldsfor ultra cold atomic systems are called "synthetic". In the next section \ref{sec:SO} as a specific example  of syntehtic non-abelian gauge field, we shall explain how synthetic spin-orbit coupling for ultra cold gases. We shall finlly conclude.

%%%%%%%%%%%%%%%%%%%%%%%%%%%%%%%%%%%

\section{Gauge Invariance of Schor\"odinger Equation in Electromagnetic Field}\label{Abl}
In this setion we shall start with the  gauge invariance properties of the ordinary Schr\"odinger Equation 
in Quantum Mechanics, namely 
how this equation transforms under the gauge transformation and how does it compare with our knowledge 
about gauge invariance from classical physics. Though any standard book on "Quantum Mechanics" may be consulted for more detail, we shall closely follow the discourse given in the book by  J. J. Sakurai \cite{JJS}

In quantum mechanics our description of a physical system starts with Schr\"odinger equation 
\beq  i \hbar \frac{\partial \Psi}{\partial t} = H \psi \nonumber \eeq 
The solution of this equation can be written 
as \beq \Psi (\bs{x}, t) = \mathcal{U}(t) \Psi (\bs{x},0) \nonumber \eeq
Where the time evolution operator $\mathcal{U}(t) = \exp (-i \frac{\hat{H} t}{\hbar})$. We shall consider the quantum mechanical motion of  a charged particle in presence of electric and magnetic field. Such a situation is very common to many 
solid state electronic systems where one can subject the free electrons that carry electric current to such 
forces. According to the law of electrodynamics such electric and magentic field can be given in terms of 
scalar potential $\phi$ and vector potential  $\bs{B}$ such that 
\beq \bs{E} = - \bs{\nabla} \Phi ; \bs{B} = \bs{\nabla} \times \bs{A} \nonumber \eeq 
For the discussion below we shall only concentrate on the vector potential alone, which can be wasily generalized 
to scalar potential. This is because according to the convention of Special Relativity, scalar and vector potential are 
just the time and space like components of a Four Potential. 
\beq \text{Since,}~\bs{\nabla} \times \bs{\nabla} \Lambda=0  \Rightarrow \bs{B} = \bs{\nabla} \times \bs{A'}= \bs{\nabla} \times (\bs{A} + \bs{\nabla}\Lambda), \label{vector} \eeq 
 
Similarly if 
\beq V'= V - \frac{1}{c} \frac{\partial \Lambda}{\partial t}~ \text{and} ~ \bs{A'} = \bs{A} + \bs{\nabla} \Lambda \nn \eeq
Then 
\beq \bs{E} = -\bs{\nabla} V - \frac{1}{c} \frac{\partial \bs{A}}{\partial t} = 
-\bs{\nabla} V' - \frac{1}{c} \frac{\partial \bs{A'}}{\partial t}   = \bs{E'}  \label{scalar} \eeq 
Therefore,  classical electrodynamics tells us that 
upto a gauge transformations the vector and scalar potentials are arbitrary. This means that 
two vector potentials differ from each other by a gradient of a well behaved scalar function  $\Lambda$, will 
lead to the same magnetic field which is clearly a  measurable quantitity. 
This non uniqueness of the vector potential for a given magnetic field however does not create a problem 
in describing the motion of charged partile in classical mechanics under Newton's Laws. This is because 
the force on such classical charged particle, Lorenz force is given by 
\beq F = q \bs{v} \times \bs{B} \nonumber \eeq. 
Therefore in classical physics the fundamental quantity is magnetic field and vector potential is more like mathematicl quantity that defines such magnetic field upto a gauge transformation. By this statement all 
the physically measurable quantities must depend on the magnetic field and not on the vector potential. 

We shall first construct the definition of gauge transformation from the preceeding discussion and then continue to improve over it in this section before going to ultra cold atoms. Summarizing the above description, such a  transofrmation is the one that links a vector 
potential with another with both yielding the same magnetic field. Magnetic field is therefore a gauge invariant 
quantity and only this appears in the classical equation of motion. This is why all measureable dynamical 
variables in particle mechanics are gauge invariant. Please note at this stage we did not mention anything to 
the field part of the Lagrangian which is required for a complete description of the problem. This is done keeping 
in mind the special situation in ultra cold atoms which we want to analyze in this particular review.
One can therefore expect at the level of Quantum mechnaics where we only quantize the motion of the partilce and use the same field as used in the classical mechanics, such gauge invariance should be respected. 

The Hamiltonian in presence of such magnetic field is given as 
\beq H = \frac{1}{2} (\bs{p} - \frac{e}{c}\bs{A})^{2} \nonumber \eeq.
In the above expression , the canonical momentum is $\bs{p}$  is distinguished from the mechanical momentum 
\beq \bs{\Pi} = \bs{p} - \frac{e}{c} \bs{A} \nonumber \eeq 
However the previously proposed idea about gauge invariance now needs a careful scrutiny since the vector potential now appears directly in the Hamiltonian.  
It can be checked that the commutation relations between different components of the mechanical momentum does not vanish, 
\beq [ \Pi_{i} , \Pi_{j} ] = \frac{i \hbar e}{c}\varepsilon_{ijk}B_{k} \nonumber \eeq 
However the commutator is indeed gauge invaraint. 
Using the above commuttator and the fact that 
\beq H = \frac{\bs{\Pi}^{2}}{2m} \nonumber \eeq 
It can be straightforwardly shown that the spectrum is given by  so called one dimensional harmonic oscillator like Landau levels 
\beq E_{n} = ( n+ \frac{1}{2} ) \hbar \omega_{c}  ; \omega_{c} = \frac{eB}{mc} \nonumber \eeq 
The energy spectrum is thus a Gauge invariant quantity. However the gauge invariant form of the energy 
and the basic commutation relation not necessarlty ensures that relevant physical quantities in quantum mechanics, such as the transition matrix elements two different states under the action of a given operator are necessarily gauge invariant. 

To understand this issue better, let us recall the Ehrenfest theorem which states that expectation values 
of the observables in quantum mechanics behave in the same way like the classical quantities. Therefore 
we can expect them to transform in the same gauge invariant way like classical quantities. 
As one can see this is not trivially staisfied, since what appears in the dynamical variable like Hamiltonian  
is $\bs{A}$ and not $\bs{B}$. This tells us that under a gauge ransformation the operators 
indeed gets afftected.
To see how the gauge invriance of expectation values can be ensured, let us define a state ket $|\alpha \rangle$
in presence of vector poetential $\bs{A}$ and the corresponding state ket $|\alpha'\rangle$ for the same magnetic 
field with a different vector potential $\bs{A'} = \bs{A} + \bs{\nabla} \Lambda$. 
Our basic requirement for gauge invariance is 
\bea  \langle \alpha |\bs{x} | \alpha \rangle & = & \langle \alpha' | \bs{x} | \alpha' \rangle \nonumber \\
\langle \alpha |(\bs{p} - \frac{e}{c}\bs{A}  | \alpha \rangle & = & \langle \alpha' | \bs{p} - \frac{e}{c}
\bs{A}'| \alpha' \rangle \eea 
apart from the normality of each ket. Now since both kets are normalized  there must be a unitary operator
such that 
\beq | \alpha' \rangle = \mathcal{G} |\alpha \rangle \label{U1WF} \eeq 
The invariance of position and momentum expectation value then demands 
\bea  \mathcal{G}^{\dagger} \bs{x} \mathcal{G} = \bs{x} \nonumber \\
\mathcal{G}^{\dagger} ( \bs{p} - \frac{e}{c} \bs{A}  - \frac{e}{c} \bs{\nabla} \Lambda ) \mathcal{G} & = & 
\bs{p} - \frac{e}{c} \bs{A} \nonumber \eea 
One can immediately see that the unitary operator that does that job is 
\beq \mathcal{G} = \exp [\frac{ie}{hc} \Lambda (\bs{r}) ] \label{U1} \eeq 
This is actually the generator of $U(1)$ gauge transformation and is same as a phase transformation. 
This is also the simplest gauge transformation. The moral of the above strory is that in Quantum Mechanics to 
to keep dynamical variables $U(1)$ gauge invariant, the wavefunction also need to changes under gauge tranformation and acquire an additional phase. 
 
This has highly non trivial consequences such as Aharanov-Bohm effect. However to stay focussed on our topic 
in the next section we shall not discuss this issue further. The function $\Lambda$ that appeared as the exponent and implenets the gauge tranformation is a function of local cordinates. Since all such functions comuutes with each other, such a gauge transformation is called Abelian. In the next section we shall consider more complicated gauge tranformations which are non-Abelian. 

\section{Non-Abelian Phases}\label{NonAb}
We shall now introduce Non Abelian gauge field using the language of quantum mechanics rather than quantum field theory. To this purpose we shall follow the treatment given in  Ref. \cite{Quantum}.  
Yang and Mills in 1954 generalized this gauge(phase) invariance properties of the Schoredinger Equation 
for multicomponent wave function, namely when wavefunction has internal degrees of freedom apart from 
the co-ordinate space or orbital degrees of freedom. As we know when the wavefunction has such internal degrees of freedom, then the wavefunction is a complex vector defined at each point in space time rather than a complex number. In such case 
\beq \bs{\Psi} (\bs{r}, t) = \begin{bmatrix} \Psi_{1}(\bs{r},t) \\ \Psi_{2}(\bs{r}, t) \\ \cdots \\ \Psi_{N-1}(\bs{r}, t) \\  \Psi_{N}
(\bs{r}, t) \end{bmatrix} \nonumber \eeq

We shall now extend the concept of Gauge invariance for a sacalar Schr\''odinger equation in the earlier section, 
to the case of vector Schor\"dinger equation that will be satisfied by such multicomponent wave function and will analyze the consequences. However unlike in the previous case, here  we pretend that initially have no idea 
what type of "Electromagnetic field" will demand the resulting gauge invarinace. So we started by demanding 
generalization of such gauge invariance for a multi-component Schr\"odinger equation and wait for the outcome.
 
SInce here the wavefunction is multicoponent the generalization of $\exp ( i \frac{e}{hc}\Lambda( x))$ will be an unitary matrix, where the condition for unitarity comes from the constraint that the norm of the wavefunction 
\beq |\bs{\Psi}(\bs{r}, t)|^{2} = |\Psi_{1}(\bs{r}, t)|^2 + |\Psi_{2}(\bs{r},t)|^{2} + \cdots +  |\Psi_{N}(\bs{r}, t)|^{2} \eeq 
should 
remain invariant under this transformation 
\beq \bs{\Psi'}(\bs{r}, t) = U \bs{\Psi}(\bs{r}, t) \nonumber \eeq 
If we apply this transformation to the Schr\"odinger equation for the multicomponent wavefunction which is 
( here we assume that all operators are correctly multiplied by suitable matrices so that they have the 
correct dimension), 
\beq i \hbar \frac{\partial \bs{\Psi}}{\partial t} = (-i \hbar \nabla - \frac{e}{c} \bs{A} )^{2} \bs{\Psi} (\bs{r}, t) + e \bs{V} \bs{\Psi} 
(\bs{r}, t), \label{nonab1} \eeq 
it can be written as 
\beq i \hbar \frac{\partial}{\partial t} U^{-1} \bs{ \Psi'}(\bs{r}, t) = (-i \hbar \bs{\nabla} - \frac{e}{c} \bs{A})^{2} U^{-1}
\bs{\Psi'}(\bs{r},t) +e\bs{V}(\bs{r}) U^{-1}\bs{\Psi}(\bs{r},t) . \nonumber \eeq 

We multiply both sides of the above equation from left by $U$ and expand the covariant momentum operator 
inside bracket and finally obtain.
\bea i \hbar \frac{\partial}{\partial t}\bs{\Psi}'(\bs{r}, t) & = & (- i \hbar \bs{\nabla} -\frac{e}{c} U \bs{A} U^{-1} - i\hbar U \nabla U^{-1})^{2} \bs{\Psi}'(\bs{r},t) \nonumber \\
&  & \mbox{} + eU \bs{V} U^{-1} \bs{\Psi}'(\bs{r},t) -i\hbar U\frac{\partial}{\partial t}U^{-1} \bs{\Psi}'(\bs{r},t) 
\label{nonab2}
\eea 
as the Schr\'odinger equation for transformed wave function $\bs{\Psi}'$. This Eq. \ref{nonab2}
will have the same form as Eq. \ref{nonab1} if we define 
\bea \bs{A'} & = & U \bs{A}U^{-1} +  i \frac{\hbar c}{e} U \nabla U^{-1}  \nonumber \\
\bs{V'} & = & U \bs{V} U^{-1} - i \frac{\hbar}{e} U \frac{\partial}{\partial t} U^{-1} \label{nonab3} \eea 
The above set of transformations define the rules for gauge tranformations for a multicomponent wave function. 
 The gauge tranformations defined in this way through the unitary matrix $U$ is a generalization of 
the one done for scalar wave function in Sec. \ref{Abl}, but now $\bs{A}$ and $V$ are matrices. To see this connetction explicitly let us note that any 
 unitary matrix $U$ can be written as 
\beq U = e^{i \mathcal{H}}  \nonumber  \eeq
where $\mathcal{H}$ is a Hermitian matrix. In case  the a scalar wavefunction, $\mathcal {H} = \frac{e}{hc} \Lambda(\bs{r})$ . Since Hermitian matrices in general do not commute the gauge fields that transforms 
according to the tranformations defined in (\ref{nonab3}) are called Non-Abelian gauge fields, whereas for 
the scalar wave function they are Abelian. 

What will be the corresponding gauge field for such non-abelian gauge potential? To find out that 
let us note , that the principle we adopted in finding out the electromagnetic field is that under the gauge tranformation 
\beq A_{\mu}(x) \rightarrow A_{\mu} + \partial_{\mu} \Lambda(x) \nn \eeq
This is basically the same equations written earlier in 
Eq. (\ref{scalar}) and Eq.(\ref{vector}), but written in more compact way using relativistic notation.  
The field strength should remain invariant under such gauge transformation. This implies that 
the field strength can be given by 
\beq F_{\\mu \nu} = \partial_{\mu} A_{\nu} - \partial_{\nu} A_{\mu} \label{AbelianField} \eeq 
It can be readily checked that under the Abelian gauge tranformation  this is indeed gauge invariant.  Also in 
(\ref{AbelianField}) the terms that only contains the spatial derivative of gauge potential, combines to give 
\beq \bs{\nabla} \times \bs{A} = \bs{B} \nn \eeq.

For Non-Abelian gauge potential under the transformation defined in \ref{nonab3} it can be shown that 
\bea F_{\mu \nu} & \rightarrow &  \partial_{\mu}[U^{-1}A_{\nu} A + i U^{-1} (\partial_{\nu} U) ] - \partial_{\nu} [U^{-1}A_{\mu} A + i U^{-1} (\partial_{\mu} U) ] \nn \\
& = & U^{-1}(\partial_{\mu} A_{\nu} - \partial_{\nu} A_{\mu} ) U \nn \\
&+ & \mbox{} i [ (\partial_{\mu}U^{-1})(\partial_{\nu} U) -(\partial_{\nu}U^{-1})(\partial_{\mu} U) \nn \\
& +& \mbox{}  (\partial_{\mu}U^{-1})A_{\nu} U  + U^{-1}A_{\nu} (\partial_{\mu} U) - (\partial_{\nu}U^{-1})A_{\mu}U
- U^{-1}A_{\mu}(\partial_{nu} U) \eea 
It is clear the same expression does not transform covariantly. To make it covariant let us note 
that under the same gauge tranformation, 
\bea -i[A_{\mu}, A_{\nu}] & \rightarrow & -i U^{-1}[A_{\mu} , A_{\nu}]U \nn \\
& & \mbox{} -i [ (\partial_{\mu}U^{-1})(\partial_{\nu} U) -(\partial_{\nu}U^{-1})(\partial_{\mu} U) \nn \\
& & \mbox{} - (\partial_{\mu}U^{-1})A_{\nu} U  - U^{-1}A_{\nu} (\partial_{\mu} U) +(\partial_{\nu}U^{-1})A_{\mu}U
+U^{-1}A_{\mu}(\partial_{nu} U) \eea 
This means the additional terms that appears in the expression of $F_{\mu \nu}$, also appears in the 
the transfomed commutator of the Non Abelian gauge potentials, however with the opposite sign.   
This suggest that if we define 
\beq F_{\mu \nu} = \partial_{\mu} A_{\nu} - \partial_{\nu} A_{\mu} -ig[A_{\mu}, A_{\nu} ]  \label{NONABF} \eeq 
then it transforms covariantly under the gauge transformation. Here the quantity $g$ depends on the 
nature of the coupling with the gauge potential. 
Equation (\ref{NONABF}) defines the Non Abelian field strength.

Non Abelian gauge fields do appear in nature and fields tranforming according to the rules given in Eq. (\ref{nonab3}) are actually responsible for weak and strong interaction that happens inside nucleus. This 
is one of the dominant topic in quantum field theory \cite{FTB, Field}. But as the above discussion emphasizes that 
they may appear in ordinary quantum mechanics also. This is what we are going to discuss further in the next section. One of the motivation for that it shares a close connection with synthetic gauge field for ultra cold atoms. 

\section{Geometric Phase in Quantum Mechanics and the related Gauge Fields} \label{sec:Berry}
Abelian and Non Abelian gauge fields are fundamental to our understanding of the nature since it is known 
three fundamental interactions Electromagnetic, Weak and Strong are due to the existence of such gauge fields.
However, the Abelian gauge transformations are equivalent to phase transofrmations and Non Abelian Gauge transformations are their higher dimensional generalization.  One can therefoee naturally ask the question  
whether such a transformation arises in the other domain of Quantum Mechanics and if the answer 
is yes what are the possible realizations. This question was answered in the clearest form by M. V. Berry 
in his seminal work \cite{Berry} based on a number of other works which already indicated the existence of such different type of gauge fields in a number of physical phenomena, that spans optics \cite{Pancharatnam}, Chemistry \cite{Mead}, Atomic and Molecular Physics \cite{Jackiw} etc. A selected collection of papers on this 
related topic is available in Shapere and Wilczek edited book \cite{Shapere} . Here we follow a  pedagogical account of the 
key argument given in Ref. \cite{Shankar}.

To this purpose we shall consider the the phase change in a quantum mechanical wavefunction under an 
adiabatic change. The adiabatic theorem in quantum mechanics tells us that if the particle Hamiltonian is given by 
$H(R(t))$ where $R$ is some external co-ordinate which changes suffiiciently slowly (slower than the natural time 
scale set by the typical enegy spacing in the unperturbed system)
and appears parametrically in $H$,  then the particle will sit in the $n$-th instantaneous eigenstate of $H (R(t))$ at the time $t$ if it started out in the $n$-th eigenstate of $H(R(0))$

The solution of the time dependent Schr\"odinger equation for this case is 
\beq |\psi(t) \rangle = c(t) \exp ( -\frac{i}{\hbar} \int_{0}^{t} E_{n}(t') dt') |n(t) \rangle) \label{Berry1} \eeq 
Here the exponential factor comes from the usual time evolution of an eigenstate of the Hamiltonian, after taking 
into account the fact that one is dealing with the instantaneous eigenstate of the time dependent Hamiltonian which is changing with time. 
The other 
factor $c(t)$ is kept to check the fact if due to the time evolution of the basis states is there any non-trivial additional time dependence.  
Substituting the  state (\ref{Berry1}) in the time dependent Schr\"odinger equation and taking the inner product with 
the instantaneous $\langle n(t) |$ one gets  
\beq  \frac{dc(t) }{dt} = -c(t) \langle n(t) | \frac{d}{dt} | n(t) \rangle \nonumber \eeq 
with the solution 
\beq c(t) = c(0) e^{i \gamma(t)} \nn \eeq 
with $\gamma(t) = i \int_{0}^{t} \langle n(t') | \frac{d}{dt'} | n(t') \rangle dt'$. 
The important thing here to notice that this phase is arising because the basis state $|n(t) \rangle$ is constantly 
changing with time. The instantaneous adiabatic state can therefore be writtes 
as 
\beq |n (R) (t) \rangle_{a} = e^{ i \gamma (t)} | n(R(t))  \rangle \nn \eeq
where the subscript $_{a}$ is used to denote the difference with a time evolved state in the absence 
of such phase factor. We know that an extra phase factor in a quantum mechanical state 
may not have any measurable consequences. However  
in the presence case actually it does have.  We shall explian it here. Let us rewrite  
\bea exp(- \int_{0}^{t} \langle n(t') | \frac{d}{dt'} | n(t') \rangle dt') & = & 
\exp( \frac{i}{\hbar} i\ int_{0}^{t} \rangle n(t') | \frac{d}{dt'} | n(t') \langle dt' ) \nn \\
&= & \exp ( \frac{i}{\hbar} \int_{0}^{t} i \hbar \langle n(R(t') |\frac{d}{dR} | n(R(t') \rangle \frac{dR}{dt'} dt') \nn \\
& = & exp (\frac{i}{\hbar}\int_{0}^{t} A^{n}(R) \frac{dR}{dt'}dt') \label{Berry} \eea 
Where \beq A^{n}(R) = i \hbar \langle n(R) | \frac{d}{dR} | n(R) \rangle \nn \eeq 
is known as the "Berry curvature". 
Now  
under a phase transformation on the state 
\bea |n(R) \rangle  &\rightarrow&  \exp ( i \Phi(R) ) | n(R) \rangle = |n'(R) \rangle \nn \\
\text{Berry Curvature transforms as}~ A^{n}(R) &\Rightarrow&  A^{n}(R) - \hbar \frac{d \Phi(R)}{dR} \label{Berry2} \eea 
This is exactly the gauge invariance condition that was imposed on the vector potential 
in Eq. (\ref{vector}) in the previous section \ref{Abl}.  The transformation of the wavefunction defined in (\ref{Berry2})
is same as the one defined in Eq. (\ref{U1WF}) for real electromagnetic field. 
What motivates a gauge invariant quantity in this case?  To see that, consider a case where the adiabatic parameter $R$ comes back to the same value after a time period $T$. 
This implies 
$R(T)=R(0)$ and $H(T)=H(0)$. Under that case the singlevaluedness of the wave function in the parameter (R)
space demands that the line integral of the Berry curvature around the closed loop in the parameter space  must 
be invariant under such gauge or phase tranformation. 
This is the same condition which states that two vector potential $\bs{A}$ and $\bs{A'}$ differing from each other through a gauge transformation when integrated over a closed contour will be same since this is equal to the flux enclosed by the area ($ \int \bs{B} \cdot d\bs{S}$). 
Thus the Berry curvature exactly plays the same role as the vector potentail due to a real
magnetic field under gauge transformation and its effect on the wave function, namely the integral of the 
vector potential around a close loop in the co-ordinate space is gauge invariant as demanded by the single 
valuedness of the wavefunction. 

A pertinent question at this point is whether such adiabatic evolution of the time (parameter) dependent Hamiltonian will also generated a scalar potential along side a vector potential. This is important to establish to
full analogy with electromagnetic theory, since vector and scalar potentials are space and time like component
of the four potential that apears in a relativistically invariant theory of Electromagnetism. It turns out in this case theire also exists a corresponding scalar potential which has the form 
\beq V(R) = \frac{\hbar^{2}}{2m}[ | \frac{d}{dR}| n(R) \rangle|^{2} - \langle \frac{d}{dR} n(R) | n(R) \rangle \langle 
n(R) | \frac{d}{dR}| n(R) \rangle \nonumber \eeq . We again refer to Ref. \cite{Shankar} for the detailed derivation of 
this accompanying scalar potential. The gauge potentials described in this section 
are known in the literature as  geometric gauge potential  because of their origin. 

The adiabatic parameter that appears here is not necessarily a scalar, it could be a vector as well, namely 
$\bs{R}$, having a certain number of components.  
In that case a straightforward generalization of the above calculation will show in that case Berry Curvature 
will be a matrix and its different component will generate the Non Abelian counterpart of the Geometric Gauge potential. 
The most significant impact of the concept of "Berry curvature" or "Geometric Vector Potential" is that it  opens the possibility of identifying gauge poetntial and fields in a wide variety of quantum systems. In the following section 
we shall analyze how these concept leads to the creation of synthetic gauge field for ultra cold atoms.

\section{Rotating Bose-Einstein Condensates}\label{RotBose}
The simplest example of implementing the artificial or synthetic gauge field for cold atomic condensates is through rotation. 
This was accomplished  by ENS Group \cite{Dalibard}, MIT group \cite{Ketterle} 
and JILA  group \cite{Cornell}. This method exploits the equivalence between the Coriolis force in a rotating frame and Lorenz force acting on an electron  in a uniform magnetic field \cite{Fast}. 
In this scheme the trap in which an ultra cold condensate is created is rotated by using a moving laser. 

\begin{figure}
\centering
\includegraphics[width=0.85\columnwidth , height= 
0.65\columnwidth]{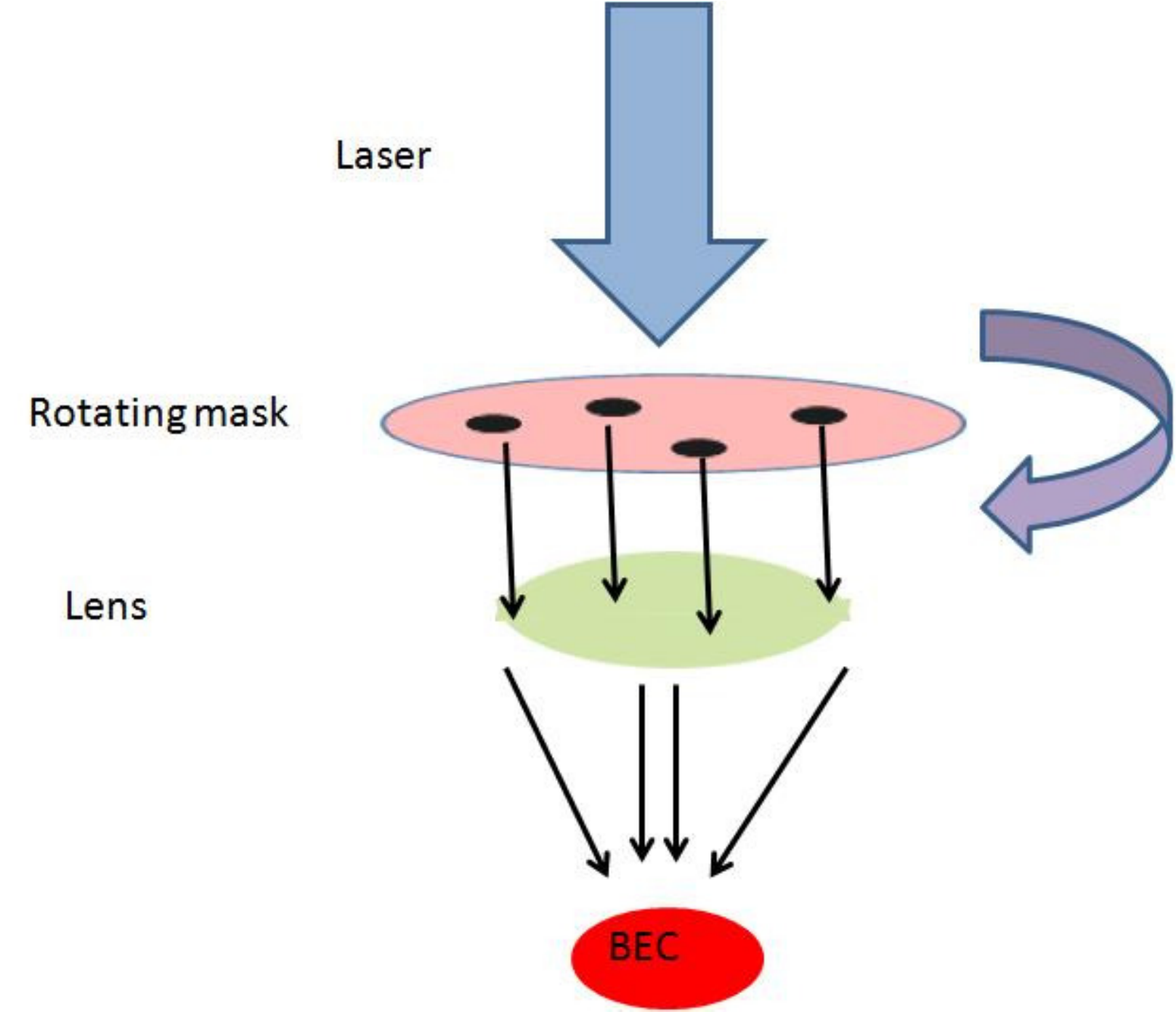} 
\caption{A typical configuration for rotated trap. Here the laser indced optical potential imprinted on BEC is rotated
with the help of a rotating mask. In some experiments such a set up was used.}
\label{rotation}
\end{figure}

If the plane 
of this rotation is taken as $x-y$ plane and the symmetry axis as $\hat{z}$, the effect  of such rotation on spatial 
co-ordinate is given by  
\beq R_{z}(\phi) \begin{bmatrix} x \\ y \end{bmatrix} R_{z}(\phi)^{\dagger} = 
\begin{bmatrix} \cos \phi & \sin \phi \\
-\sin \phi & \cos \phi \end{bmatrix}  \begin{bmatrix} x \\ y \end{bmatrix} \nn \eeq 
Here $R_{z}( \phi) = exp(-i \frac{\phi \hat{L}_{z}}{\hbar})$ is the Rotation operator about $\hat{z}$ axis.
If the rotation is executed at an uniform angular velocity $\Omega$, then $ \phi = \Omega t $. We can immediately see the connecion between $R_{z}(\phi)$  and the U(1) gauge transformation defined in (\ref{U1}). 

The time dependent Hamiltonian that describes a trapped boson in a rotating frame is given 
by 
\beq H(t) = R_{z}(\Omega t) [ \frac{\bs{p}^{2}}{2m} + \frac{1}{2}m (\omega_{x}^{2} x^{2} + 
\omega_{y}^{2} y^{2} )] R_{z}^{\dagger} (\Omega t) \nn \eeq 
Since the $\bs{p}^{2}$ remains invariant under the rotation it yields 
\bea H(t) & = & \frac{\bs{p}^{2}}{2m} + \frac{m}{2}[(\omega_{x}^{2}(x \cos \Omega t + y \sin \Omega t)^{2} \nn \\
 & + & \omega_{y}^{2} ( -x \sin \Omega t + y \cos \Omega t)^{2} ]  \label{rotham} \eea  

One neeeds to solve the  time dependent Schr\"odinger equation (TDSE) for such system which is 
\beq i \hbar \frac{\partial \psi}{\partial t} = H(t) \psi. \label{TDSE} \eeq
However to do equilibrium thermodynamics of a systems of such bosons it it useful to go to the co-rotating from 
where the Hamiltonian does not change with time.  For that 
one needs to do a unitary tranformation on the wave function by writing 
\beq \psi ' = R_{z}^{\dagger} (\Omega t) \psi \nn \eeq 
This is again equivalent to the gauge transformation given in (\ref{U1WF})
where 
the $\psi '$ is the wave function in the co-rotating frame. 
The transformed  TDSE for $\psi '$ looks like 
\beq i \hbar \frac{\partial \psi'}{\partial t} = [ \frac{\bs{p}^{2}}{2m} + \frac{m}{2} (\omega_{x}^{2} x^{2} + \omega_{y}^{2} y^{2} )  + \Omega L_{z} ] \psi'  \label{TDSE1} \eeq

One can check the time independent Hamiltonian on the right hand side can be written 
as 
\beq H = \frac{ (\bs{p} - m \bs{A})^{2}}{2m}  + \frac{1}{2}m [ \omega_{x}^{2} x^{2} + \omega_{y}^{2} y^{2} - \Omega^{2} r^{2} ]  \eeq 
Thus the unitary transformation that takes the wavefunction to the co-rotating frame, also induces a gauge potential in the stationary hamiltonian in the co-rotating frame. Comparing with our discussion in section 
\ref{Abl} we can comment that the unitary operator $R_{z}( \phi)$ defines here 
is mathematically same as unitary gauge transformation operator $U= \exp ( i \frac{e}{hc} \Lambda(x))$ defined  The tranformation of the Hamiltonian operator introduces a gauge potential. 

The gauge (vector) potential and the gauge field obtaained in this way is of the form 
\bea \bs{A} & =  & - \Omega y \hat{x} + \Omega x \hat{y} \nn \\
         \bs{B} & = & 2 \Omega \hat{z} \nonumber \eea

This is however not the only effect on the Hamiltonian by the unitary tranformation. It will also introduce a scalar potential 
\beq  V_{R} (\bs{r}) = -\frac{1}{2} m \Omega^{2} r^{2} \label{decon} \eeq 
Thus the effective trap potential in the rotating frame gets reduced. 

As we can see we can equivalently the creation of such "artificial" gauge field through rotation can also be explained by using the concept of Berry Curvature discussed in section \cite{Berry}. One can recognize here 
that the adiabatic parameter is the time dependent rotation angle $\Omega t$ and the Hamiltonian is a function 
of this parameter $R=\Omega t$. If the rotational frequency is ramped up adiabatically, then it can be ensured 
that system will always stay in the ground state of the rotated hamiltonian provided the initial system is in the ground stat

\begin{figure}
\centering
\includegraphics[width=0.85\columnwidth , height= 
0.65\columnwidth]{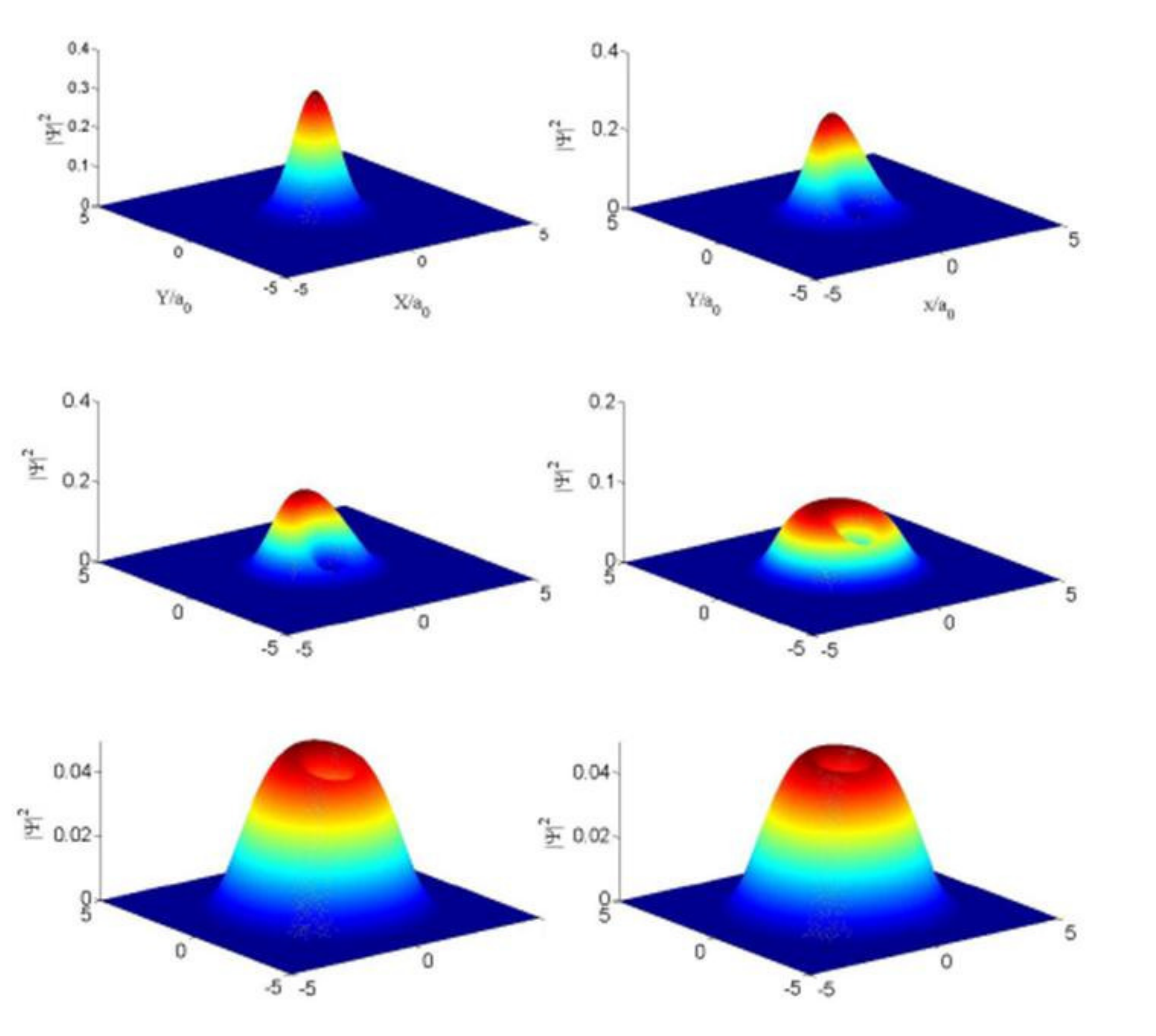} 
\caption{The entry of a vortex in a rotating condensate with increased rotational frequency and the change of
condensate profile. The sequence of the plots will be like $ (1,1) \rightarrow (1,2) \rightarrow (2,1) \rightarrow (2,2)
\rightarrow (3,1) \rightarrow (3,2)$.
The numerical result is obatined by simulating Gross-Pitaevskii equation (\ref{GP}) in a co-rotating frame.
The $x$ and $y$ axis is the condensate co-ordinates in dimensionless units. We also take $\omega_{x}=\omega_{y}=\omega_{\perp}$. The $z$ axis is the 
superfluid density $|\Psi^{2}$ At the position of the vortex the complex order parameter of the superfluid condensate $\psi$ vanishes and the phase of complex 
order winds around the position of the vortex.}
\label{vortex}
\end{figure}

The ultra cold atomic Bose Einstein condensate we are going to talk about consists of interacting bosons. However 
in the typical experimental condition, the system is described very well by the mean field Gross-Pitaevskii equation 
\cite{Stringari}. For simplicity we choose a two dimensional condensate that can be derived from a three dimensional model for 
a rotating cigar shaped condensate. The Gross-Pitaevskii equation is 
\beq i \hbar \frac{\partial \Psi}{\partial t} = ( -\frac{\hbar^{2}}{2m} \nabla^{2} + \frac{1}{2} m (\omega_{x}^{2} x^{2} 
+ \omega_{y}^{2} y^{2} ) + g |\Psi |^{2}) \Psi \label{GP}  \eeq
Even though the $\Psi$ appeared in the above equattion is the mean field superfluid order parameter of the $N$-boson 
condensate and not the quantum mechanical wavefunction of the corresponding many body Schr\"odinger equation, the above equation has the same mathatical structure of a usual one particle Schr\"odinger equation apart from the non-linear term which represents interaction \cite{Stringari}.
It can be readily verified that the non-linear term is invrainat under the action of the Unitary operator  $R_{z}(\Omega t)$. Thus the entire previous discussion on the artificial gauge transformation of single boson Schr\"odinger equation can be applied here for the Gross-Pitaevskii equation also. In the 
early days of BEC this was the technique through which vortices and vortex lattice was created in ultra cold condensate. The entry of such vortices in a rotating condensates and the change of condensate profile due to
this is shown in the Fig. \ref{vortex}. 

We shall not discuss the vortex physics in Ultra Cold BEC any further in this review since this was 
already discussed in a number of earlier reviews. The early experiments in rotating ultra cold gases, vortices 
etc. was discussed  in \cite{rotbec1, rotbec2}. An interesting regime is where the rotational frequency $\Omega$ 
is almost equal to the trap frequency in the transverse plane $\omega_{\perp} (\sim \omega_{x,y})$. This means the the trap potential 
almost becomes negligible. 
Because of the entry of the large number of vortices in the ultra cold condensate under this condition, a number of interesting phases of large 
number of vortices appear in this regime. This is the regime of rapidly rotating ultra cold gas and have been reviewed  extensively in \cite{rotbec3, rotbec4}.

\section{The creation of Abelian and Non-Abelian Gauge Field for Ultra Cold Gases using Berry Curvature}\label{ColdBerry}
In the previous section  we have seen how synthetic Abelian field can be created for ultra cold atoms exploiting the 
similarity between the rotational operator about a particular axis with a $U(1)$ gauge transformation. 
The above scheme has a serious limitation . The additional deconfining potential in Eq. (\ref{decon}) it creates  
destabilizes the trap in which condensate is created beyond a critical 
value of the rotational frequency, when $\Omega \rightarrow \omega_{\perp}$. This in turn limits the strength of the Abelian field that can be created in this 
method. Thus the cold atom analogue of strongly correlated  phases of  two dimensional electronic systems  in a perpendicular 
magnetic field with high value ( of the order of several Tesla) such as Quantum Hall phases \cite{QH} can not be created in this set-up. This requires one to look for alternative scheme. We shall describe that in a very general way following the excellent review article \cite{dalibard} where one may look for further details. 

\subsection{Geometrically Induced Abelian Gauge Field} \label{Abel}

\begin{figure}
\centering
\includegraphics[width=0.85\columnwidth , height= 
0.65\columnwidth]{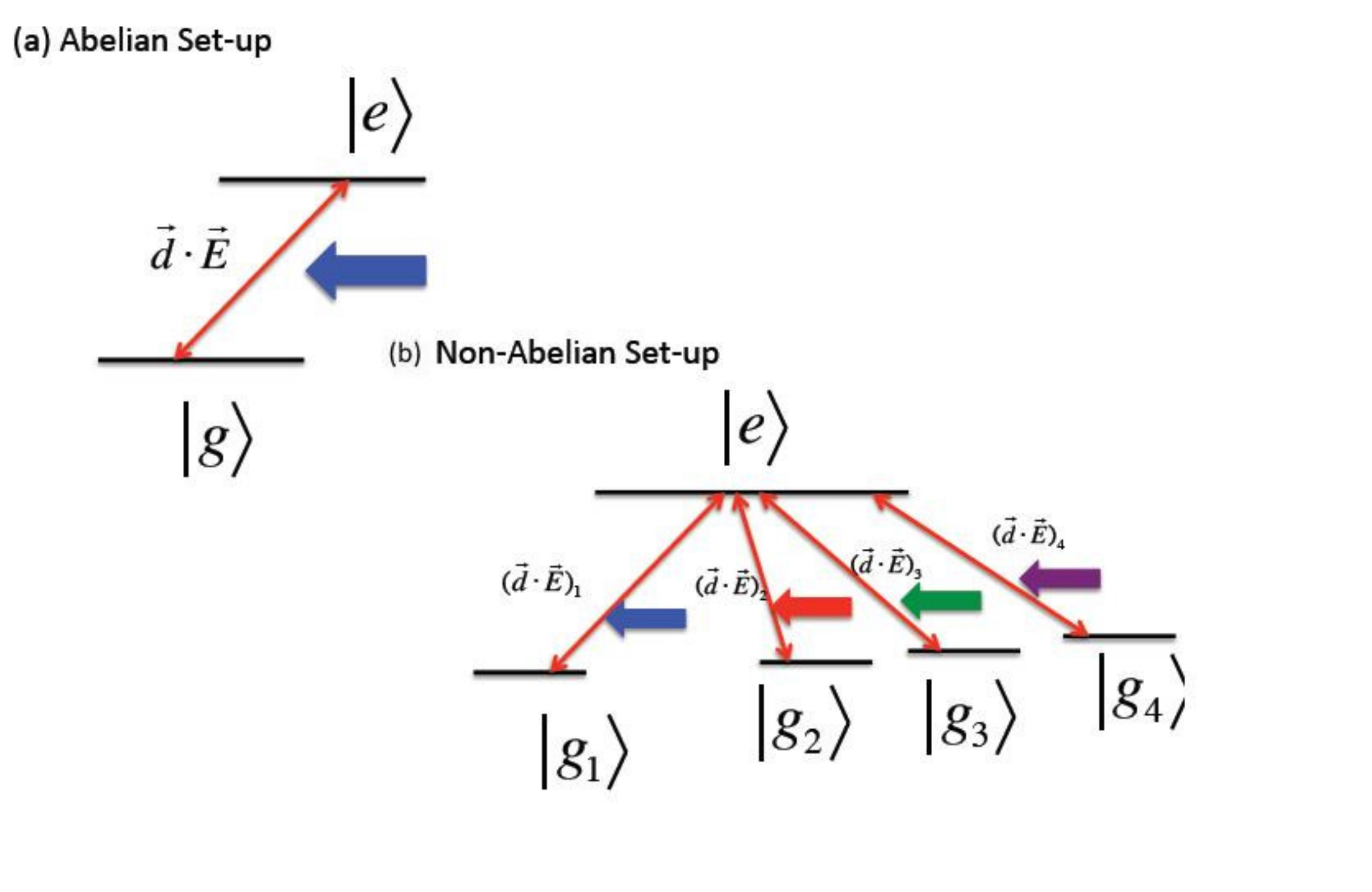} 
\caption{  (a) A typical atom-laser configuration that can be used to generate Berry curvature 
of the form of Abelian gauge potential. The atom is modelled as two level system having states $|g \rangle$ and 
$| e \rangle$. For more details refer to the discussion in Sec. \ref{Abel}
(b) Generalization of the set up  in (a) to produce Non Abelian gauge potential. For more details 
refer to the discussion in Sec. \ref{NonAbel}
}
\label{AtomLaser}
\end{figure}

Consider  a general model of a two level atom with $|g \rangle$ and $|e \rangle$ states being respectively its ground state and excited state and forms a two dimensional Hilbert space. They can be considered as the eigenstate of a simple Hamiltonian like 
\beq H_{0}= \frac{\bs{P}^{2}}{2m}  \eeq 
We consider the dynamics of the particle in  space dependent external field that couples these two states 
( Fig. \ref{AtomLaser} (a)). 
One can recognize that a given laser with 
suitable parameters can accomplish this job through dipole interaction. For more details on laser-atom 
interaction we refer to standar textbook on Quantum Optics  such as ref. \cite{QO}. 
The general Hamiltonian of  such coupled system can be written as 
\beq H_{I} = H_{gg}(\bs{r}) |g \rangle \langle g| + H_{ee}(\bs{r}) | e \rangle \langle e | + H_{ge} |g \rangle \langle e | + H_{eg} |e \rangle \langle g | \eeq
Since this is a two level system 
one can map this to spin $\frac{1}{2}$ system and rewrite this as 
\beq H_{I} = \frac{\hbar \Omega}{2} \bs{n}  \cdot \bs{\sigma}  \nn \eeq 
where $\bs{n}$ is a three dimensional unit vector parametrized in terms of polar angle $\theta(\bs{r})$ and azimuthal angle $\phi(\bs{r})$. As one can see the spatial dependence comes from the fact that the coupling between the states is assumed to be spatially dependent since it will depend on electric field of the laser and the atomic wavefunction.  

At  the spatial point $\bs{r}$, the local eigenstates of $H_{I}$ are now given by 
\bea | n_{\uparrow} (\bs{r}) \rangle & = & \begin{bmatrix} \cos \frac{\theta(\bs{r})}{2} \\
\sin \frac{\theta(\bs{r})}{2} e^{i \phi(\bs{r})} \end{bmatrix} \nn \\
|n_{\downarrow} (\bs{r}) \rangle & = & \begin{bmatrix} 
-\sin \frac{\theta(\bs{r})}{2} e^{-i \phi(\bs{r})} \\  \cos \frac{\theta(\bs{r})}{2} 
\end{bmatrix} \eea 

These two states forms a local basis for the Hilbert space at each point in the co-ordinate space $\bs{r}$.
In the language of quantum optics they are called dressed states. 
If the system evolves adiabatically through this space then this means that this local basis of the 
Hilbert space is also changing at every point in space. Following our discussion in section \ref{sec:Berry} 
this adiabatic motion will generate Berry curvature. 
 Also the quatity $i \langle n_{\uparrow} | \nabla n_{\downarrow} \rangle$ is real since $n_{\uparrow, \downarrow}$ forms an orthogonal basis. A general state 
in this Hilbert space at any point of time can be written as 
\beq |\Psi (\bs{r}, t) \rangle = \psi_{\uparrow} (\bs{r}, t) |n_{\uparrow} (\bs{r}) \rangle + \psi_{\downarrow} (\bs{r}, t) |n_{\downarrow} (\bs{r}) \rangle \label{wf}  \eeq 
Since the basis vector is changing from one point to another in co-ordinate space 
therefore 
\beq \bs{\nabla} (\psi_{i}(\bs{r}) |n_{i}(\bs{r}) \rangle) = \bs{\nabla}\psi_{i}(\bs{r}))|n_{i}(\bs{r}) + \psi_{i} (\bs{r})
|\bs{\nabla} n_{i}(\bs{r})\rangle, i= \uparrow, \downarrow \nn \eeq
Given this relation when such state is operated by the momentum operator, one yields 
\beq \bs{P} |\Psi \rangle= \sum_{i,j = \uparrow}^{\downarrow} (\delta_{i,j} \bs{P} - A_{ij})\psi_{j} |n_{i} \rangle \nn \eeq
Now suppose that an initial state the particle is in the state $ | n_{\downarrow} \rangle$ and the motional state is such that it stays in this state all the time ( the transition amplitude to the up-state is negligible). Under this condition we assume $\psi_{\uparrow}=0$ and project the Schr\"odinger equation in the dressed state $|n(\bs{r})_{\downarrow}) \rangle$. This gives us the following "gauged" Schoredinger equation for $\psi_{\downarrow}$. 
\beq i \hbar \frac{\partial \psi_{1}}{\partial t} = [ \frac{(\bs{P} - \bs{A})^{2}}{2m} + \frac{\hbar \Omega}{2} + V]\psi_{1}
\label{SchAbelian} \eeq 
Here $\bs{A}$ and $V$ are essentially the vector potential and scalar potential that arises due to the geometric 
phaase created by the the slowly changing Hilber space basis from one point to other.
The resulting vector potential is given by in this case 
\beq \bs{A} (\bs{r}) = i \hbar \langle n_{\downarrow} (\bs{r}) |\bs{\nabla} n_{\downarrow} (\bs{r}) \rangle 
=\frac{\hbar}{2} ( \cos \theta -1) \bs{\nabla} \phi \nn \eeq 

The synthetic magnetic field that is created due to the vector potential is given by 
\beq \bs{B} (\bs{r}) = \frac{\hbar}{2} \bs{\nabla} \cos \theta  \times \bs{\nabla}\phi \label{syntheticB} \eeq 
The magnetic field and vector potential crreated in this way have geometric origin.
It can concluded from the preceeding discussion on Berry's phase that this gauge potential cannot be "gauged away" completely if the magnetic field (\ref{syntheticB}) is non zero since the line integral of this vector 
potential around a closed contour in a region where the magnetic field is non-zero should be equal to the flux enclosed by this region. 
This apart, the changing basis of the Hilbert space also introduces a scalar potential 
\beq V(\bs{r}) = \frac{\hbar^{2}}{2m} |\langle n_{\uparrow}(\bs{r}) | \bs{\nabla} n_{\downarrow} \rangle |^{2} 
= \frac{\hbar^{2}}{2m} [ (\bs{\nabla} \theta)^{2} + \sin^2 \theta (\bs{\nabla} \phi)^{2} \nn \eeq 
One of the practical advantage of generating a synthetic vector and scalar potential in this way, that if we consider 
the Hamiltonian of a trapped system, the scalar potential does not offset the trap potential. The scalar potential 
in this case also can be repusive and attractive \cite{Spielman}. 
Also given the fact 
that there are various way of coupling different hyperfine states of ultra cold atomic multiplets, such a scheme 
provides one an wide range of possibilities to create such geometrically induced synthetic gauge potential and gauge field. But as we see in the next section that one of the most interesting aspect of this scheme is the fact that it can be easily generalized to create a Non Abelian-gauge field.  

\subsection{Geometrically Induced  Non-Abelian gauge field} \label{NonAbel}
One of the earliest paper that introduces the concept of such non-abelian Geometric phases in a general context iis the work by 
Wilczek and Zee \cite{Wilczek}. Here we shall discuss a specific example involving ultra cold atoms and laser
following \cite{dalibard}. 
Particularly we shall show how  idea discussed in the preceeding section can be easily generalized to creat synthetic non-abelian gauge field 
when a $N+1$ state atomic system with $N \ge 3$ is suitable configuration of laser beams. 
A prototype 
configuration is displayed in Fig. \ref{AtomLaser} (b). 
The structure of the the coupling matrix $U(\bs{r})$ will be 
\beq 
U(\bs{r}) = \begin{bmatrix} \langle 1 | U(\bs{r} ) | 1 \rangle & \langle 1 | U(\bs{r}))  2 \rangle & \cdots &  \langle 1 | U(\bs{r}) | N+1 \rangle \\
\langle 2 |U(\bs{r}) | 1 \rangle & \cdots & \cdots & \langle 2 | U(\bs{r})) | N+1 \rangle  \\

\vdots & \vdots & \vdots & \vdots \\
\langle N+1 | U(\bs{r}) | 1 \rangle & \cdots & \cdots & \langle N+1 | U (\bs{r}) | N+1 \rangle \end{bmatrix} \eeq 

For a fixed position $\bs{r}$ the above matrix can be diagonalized to give $N+1$ dressed states $|n_{i}(\bs{r}) \rangle$ with energy eigenvalues $E_{i} (\bs{r})$ where $i$ goes from $1$ to $N+1$. Under certain circumstances 
it happens that a subset $Q$ out of this $N+1$ states are either degenrate or quasi-denerate and are well separated from he rest of states energetically. It is under this candition it is possible to realize adiabatic motion 
in this low lying dgenerate subspace $\mathcal{H}_{Q}$ of dimension $Q$. Assuming that the motional states 
are such that there is almost no scattering from this low ebergy subspace $\mathcal{H}_{Q}$ to $(N+1)-Q$ higher energy state. 

Again we can write the full wave function of the 
\beq |\Psi \rangle  = \sum_{i=1}^{N+1} \psi_{i} (\bs{r}) | n_{i}(\bs{r}) \rangle \nn \eeq 
And then we can project this Schr\"oedinger equation to the reduce Hilbert space $\mathcal{H}_{Q}$ to get an equation for the reduced spinorial wavefunction $\Psi_{Q}= (\psi_{1}, \cdots, \psi_{Q})^{T}$. 
We can straightforwardedly extent the gauged Schr\"odinger   equation given in  Eq. (\ref{SchAbelian}) to its spinorial countepart, namely 
\beq i \hbar \frac{\partial \Psi_{Q}}{\partial t} = 
[\frac{(\bs{P} - \bs{A})^{2}}{2m} + \epsilon + V]\Psi_{Q} \eeq  
With the important differences that $\bs{A}$ and $V$ are now matrices with their matrix elements given by. 
\bea \bs{A}_{i,j}  & = & i\hbar \langle n_{i}(\bs{r}) | \bs{\nabla} n_{j} (\bs{r}) \rangle \nn \\
V_{i,j} & = & \frac{1}{2m} \sum_{l=Q+1}^{N+1} \bs{A} _{il} \cdot \bs{A}_{l,j} \eea
Since different component ( $x,y,z$) these effective vector potentials being matrices will not generally commute 
with each other and are therefore called non ableian vector potential.Here $\epsilon$ corresponds to the energy of the unperturbed atomic systems.

The above described atom-light interaction induced  syntehtic abelian or non-abelian gauge potential 
has been successfully implemented by I. B. Spielman's group \cite{lin1,lin2} in NIST by coupling atomic 
states with Raman lasers.
They created synthetic magnetic field, electric field as well as SO coupling in ultracold atomic systems. 
However it may be noted inspite of the fact that NIST method was able to overcome some of the difficulties 
that was encountered in rotating an ultra cold atomic system, particularly in the process of creating high "synthetic" magnetic field, it has its own limitations. Here the highest possible value of the synthetic magnetic field is capped by the wavevector of the Raman Laser. 
A detailed discussion on these experiments appeared in ref. \cite{Spielman} which one can see for more details. 
Very recently there has been experimental success in creating optical flux lattices \cite{Bloch} where it is possible to create much higher value of synthetic magnetic field. We are not covering this topic here and direct the reader 
to the refs. \cite{Cooper, NatGauge} and the refs. cited there for the same. Studying the effect of artificial gauge 
field in presence of optical lattice is another interesting topic which is also not covered in this article. A summary of 
the relevant work and discussion on some relevant issues for this topic is available in ref. \cite{Rashi}. 
The other case of synthetic gauge field for cold atoms that we shall discuss in some detail in subsequent section 
\ref{sec:SO} is the principle of creating synthetic spin-orbit coupling for ultra cold bosonic atoms. However before that 
here we shall provide a comprehensive analysis of the "synthetic-ness" of the gauge field 
created from the geometric phases. 

\subsection{Geometrical Interpretation of the Gauge Potential: Parallel Transport}\label{parallel}
In this section we shall briefly digress the origin of gauge field in Quantum Mechanics/Field Theory and discuss
the existence of similar reason in the current case of synthetic gauge field for ultra cold atoms. 
We know from quantum mechanics \cite{JJS, Shankar} that a spinorial ( two-component) wave function tranforms 
under a spin rotation as 
\beq \Psi' = \exp (\frac{i}{2}  \bs{\sigma} \cdot \bs{n} \phi ) \Psi \nn \eeq 
Here 
\beq [\sigma_{i}, \sigma_{j} ] = \epsilon_{ijk}\sigma_{k} \nn \eeq 
obeys the standard commutation relation between the the generators of the SU(2) rotation in the 
spin-spce. Under such general SU(2) transformation a $n$-dimensional iso-spinor similarly tranforms 
as 
\beq \Psi'= \exp( i M^{\mu} \Lambda^{\mu}) \Psi = \mathcal{U}(x) \Psi(x) \nn \eeq. 
In field theory a system is said to have gauge invariance, if under such tranformation the defining Lagrangian density remains 
invariant. The Lagrangian density involves the derivative of the field. Same is true for the wavefunction in Quantum Mechanics where the Hamiltonian involves the derivative of the wavefunction. 
For that it is important that the $\partial_{\mu} \Psi$, must change covariantly. Now it can be immediately 
checked that this is not the case for the usual derivatiove as 
\beq \partial_{mu} \Psi' = \mathcal{U}(\partial_{\mu} \Psi) + (\partial_{\mu} \mathcal{U})\Psi \nn \eeq
This happens because because under the generalized spin-rotation, the axes in the space of such isospin is getting at each point in space. Thus $\Psi(x)$ and $\Psi(x+dx)$ are measured in different co-ordinate system. 
To make the derivative co-variant one should compare $\Psi(x + dx)$ with the modified value of $\Psi(x)$ if it were 
transported from $x$ to $x+dx$ keeping the iso-spin axis fixed. This is known as parallel transport in isospace.
This is depicted in Fig. \ref{parallel}.
Under this condition the change in the $\Psi$ will be different and this change $\delta \Psi$ can be written as 
\beq \delta \Psi = ig M^{a} A^{a}_{\mu} dx^{\mu} \Psi \nonumber \eeq 
Here $A^{a}_{\mu}$ which takes care of the change in the local co-ordinate axes in the isospace from one point to another. The derivative defines in this way  becomes 
\bea D\Psi & = & (\Psi + d \Psi) - ( \Psi + \delta \Psi)   \nn \\
                  & = & d \psi - ig M^{a} A^{a}_{\mu} dx^{\mu} \Psi \nn \\
         \frac{D \Psi}{dx^{\mu}} = D_{\mu} \Psi &= & (\partial_{\mu} - ig M^{a} A^{a}_{\mu}) \Psi \eea

\begin{figure}
\centering
\includegraphics[width=0.85\columnwidth , height= 
0.65\columnwidth]{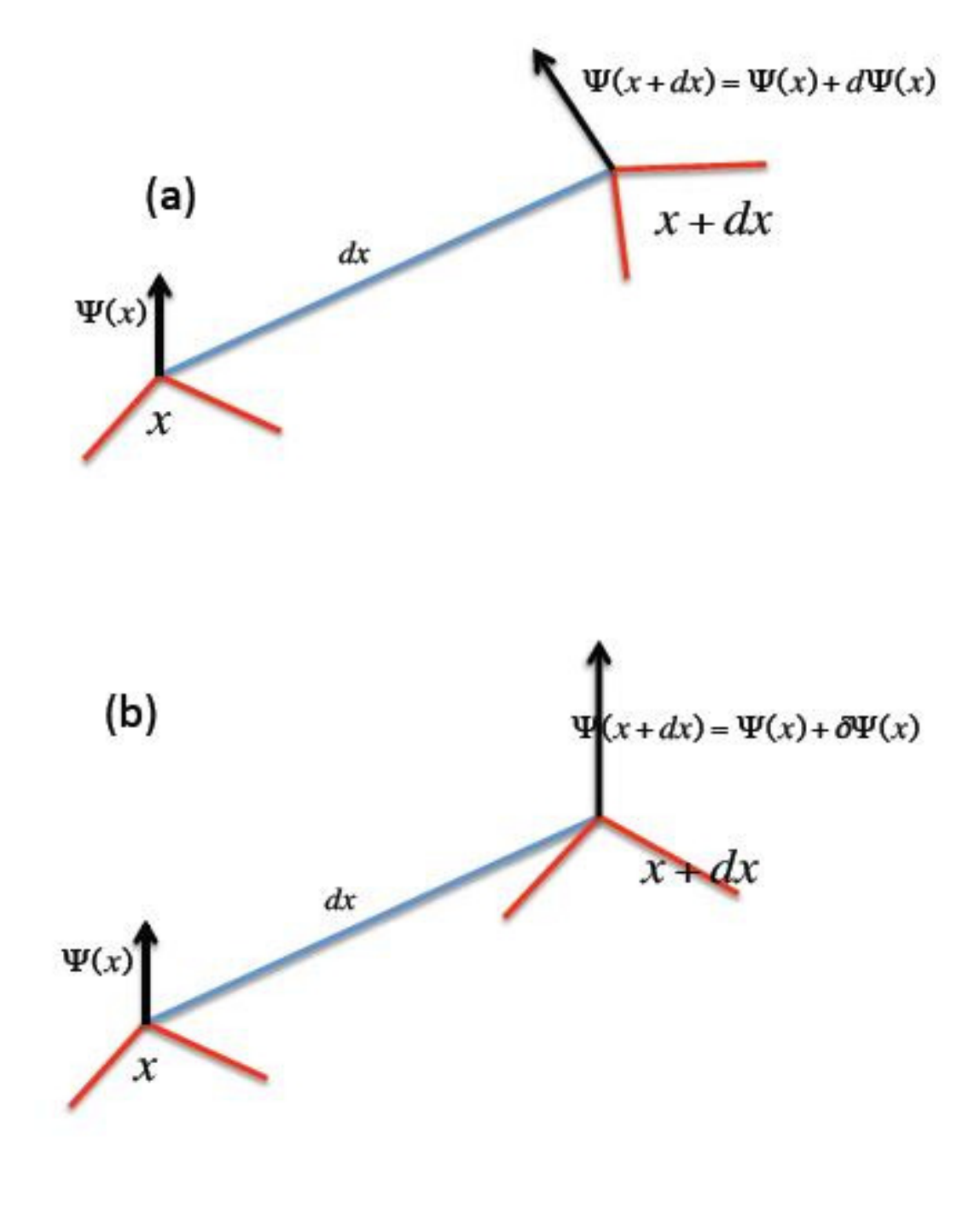} 
\caption{ The concept of parallel transport is illustrated with the help of two figures adopted from ref. \cite{Field}.
In (a) no parallel transport was done. In (b) the parallel transport was done.}
\label{parallel}
\end{figure}
It can be checked the above modified derivative transform covariantly under the guage transformation 
and is therefore qualified to enter into the  the gauge invariant Lagrangian density. Here $M^{a}$ are the generators of the rotation in the iso-spin space and their detail form depends on relevant group that represents the symmetry. The simplest of this case 
is $U(1)$ rotation where $M=1$. If particularly it corresponds to our problem of a charged particle in electromagnetic field $g=e$. Under that situation
\beq D_{\mu} = \partial_{\mu} + ie A_{\mu} \nn \eeq . 
The above discussion is available in a number of  field Theory books that discuss gauge field theory. We here mostly followed the notation and discussion of  Ref. \cite{Field}. Now comparing this discussion with the discussion in the preceeding section ( \ref{Abel} and \ref{NonAbel}), we  can immediately recognise that there also it is the chaging basis in the pseudospin space under adiabatic evolution. The adiabatic evolution of these basis states  leads to the development of syntehtic gauge field or Berry Curvature and has a similar geometrical interpretation like the true gauge field. 

Finally this brings us to a important question, namely why inspite of this similar geometrical origin, we call the Berry curvature related gauge field in ultra cold atoms as "synthetic". The reason is in true gauge field theory there 
is purely gauge field dependent term in the Lagrangian density \cite{FTB, Field} which stands for the Field 
energy. It is this term which gives the gauge field their independent dynamics in the complete matter-field Lagrangian density. Thus such gauge field are dynamical. The corresponding gauge potentials appeared in the 
covariant momentum operator. However in the atom-laser configuration the full Lagrangian density does 
not contain any such field energy term which is related to the Berry curvature generated due to adiabatic motion. 
The field part that apeears here is just the electromagnetic field energy associated with the laser and the not 
one related to the geometric gauge field. This is  why geometric gauge field do not have any independent dynamics.

\section{Synthetic Spin-Orbit Coupling for Ultra Cold Atomic Gases: Case of Non Abelian Gauge Field} \label{sec:SO}
The motivation behind creating synthetic spin-orbit coupling for ultra cold atoms primarily comes from 
the fact that spin orbit coupling plays a very important role in spinotronics \cite{Sinova} and Topological Insulators 
\cite{TI} either of which have interesting practical applications. However spin-orbit coupling also forms an interesting example for Non Abelian gauge potential which we shall describe in the following discussion.

\subsection{Non Abelian Gauge Potential and Spin-Orbit Coupling}
In our familiar notation a Non Abelian vector potential can be written as 
\beq \bs{A} = A_{x} \hat{x} + A_{y} \hat{y} + A_{z} \hat{z} \nn \eeq 
Where $A_{x}$, $A_{y}$ and $A_{z}$ are now matrices. Field strength for such 
Non Abelian vector potential given by the expression (\ref{NONABF}) can be written as 
\beq \bs{B} = \bs{\nabla} \times \bs{A} - \frac{i}{\hbar} \bs{A} \times \bs{A}  \label{NAF} \eeq
One can now easily identify that in the expression (\ref{NAF}), the first part is a straight forward 
generalization of the relation between vector potential and magnetic field for the Abelian case, the second 
part is only non zero if the gauge potential is Non Abelian. For Abelian cases, the second part is identically 
zero.

A major motivation of simulating synthetic magnetic field for ultra cold atoms is to observe Quantum Hall Effect 
like phenomena which occurs when a two dimensional electron gas is subjected to an transverse uniform 
magnetic field \cite{QH}. A related question can therefore be asked what are the gauge potentials that can create an uniform 
Non Abelian magnetic field. A detailed analysis of this probelm was done in Ref. \cite{NA}. Here we provide 
a brief summary of the relevant results to discuss subsequently in some detail  why and how one creates synthetic 
spin-orbit  coupling for ultra cold atomic systems. 

An important difference between Abelian and Non Abelian 
Gauge field is that, where as in the case of the former two vector potentials that creates the same magnetic field 
are related to each other by a simple gauge transformations, in the case of Non Abelian field that is not the case.
By that one means that two non-equivalent Non-Abelian gauge potential can lead to the same Non-Abelian magnetic field. Following \cite{NA} we shall illustrate this case for the non abelian magnetic field 
\beq \bs{B} = 2 \sigma_{z}  \hat{z} = 2 \begin{bmatrix} \hat{z} & 0 \\ 0 & -\hat{z} \end{bmatrix}  \eeq 
The above magnetic field is uniform but its direction is opposite for spin-up and spin-down component 
of the wavefunction of the particle on which it is applied. 

One type of vector potential that can give such uniform field is given by 
\beq \bs{A} = \frac{1}{2} \bs{B} \times \bs{r} = y\sigma_{z}\hat{x} -x \sigma_{z} \hat{y} \eeq
This is again a starightforward generalization of vector potential in symmetric gauge 
for uniform magnetic field $\bs{B}=B\hat{z}$ and here the vector potential contributes to the 
magnetic field only through the first term (on R.H.S.) of the  expression (\ref{NAF}) for Non Abelian 
field strength. Even though all
component of the vector potential are matrices, they are Abelian matrices. Thus this is also
the case of Abelian gauge field. The single particle spectrum of Schr\"odinger Equation 
in presence of such gauge potential  is a generalization of the Landau problem. Quantity like magnetic 
length,  phenomenon like Aharonov Bohm effect etc. can be defined for such problem. 

Another type of vector potential that can also generate the same magnetic field is given by 
\beq \bs{A} = -\sigma_{y} \hat{x} + \sigma_{x} \hat{y} \label{SONA} \eeq.
This is an uniform ( does not depend on local co-ordinate) non commuting vector potential.
This is indeed Non Abelian gauge potential. The contribution to the field purely comes from 
the second term in the expression (\ref{NAF}). The single particle spectrum of Schr\"odinger Equation 
in presence of  such gauge potential is very different from the Landau problem \cite{NA}. However there
is more interesting motivation for realizing such Non Abelian gauge potential for ultra cold atoms. 

To see this  let is recall the well known spin-orbit coupling (Thomas Term) 
which arises due to relativistic correction to the motion of a spin-1/2 electron obeying Schr\"odinger Equation, 
namely 
\beq H_{SO} = - \frac{e \hbar}{4m_{e}^{2}c^{2}} \bs{\sigma} \cdot (\bs{E} \times \bs{p}) \label{SO} \eeq

Using the tripple product  rule the above hamiltonian can be rewritten as 
\beq H_{SO} = \frac{e \hbar}{4m_{e}^{2}c^{2}} \bs{p} \cdot ( \bs{E} \times \bs{\sigma}) \nn \eeq

For a uniform electric field along $z$-axis, $\bs{E} \times \bs{\sigma}$ is just the Non Abelian uniform vector 
potential defined in Eq. (\ref{SONA}). Identifying this we can rewrite 
\beq H_{SO}  \propto (\bs{p} \cdot \bs{A}) \nn \eeq 
where $\bs{A}$ corresponds to the vector potential defined in Eq. (\ref{SONA}).
Such a term therefore also appear in the kinetic energy term of the Hamiltonian,  
\beq H_{k} =  \frac{1}{2m}(\bs{p} -m  \bs{A})^{2} \nn \eeq 
that describes a free particle in the  presence of Non Abelian gauge field (\ref{SONA}). Thus the simulation of such synthetic Non Abelian gauge field for ultra cold atoms is equivalent to create synthetic spin-orbit (SO) coupling for such systems. 
SO coupling plays a crucial role in Spinotronics \cite{Sinova} and Topological Insulator \cite{TI}.
With this background we shall now briefly discuss how such SO coupling is created experimentally for 
ultra cold BEC. More 

\subsection{Principle of Spin Orbit Coupling in Ultra Cold Bosonic Systems: NIST Method}

\begin{figure}
\centering
\includegraphics[width=0.85\columnwidth , height= 
0.65\columnwidth]{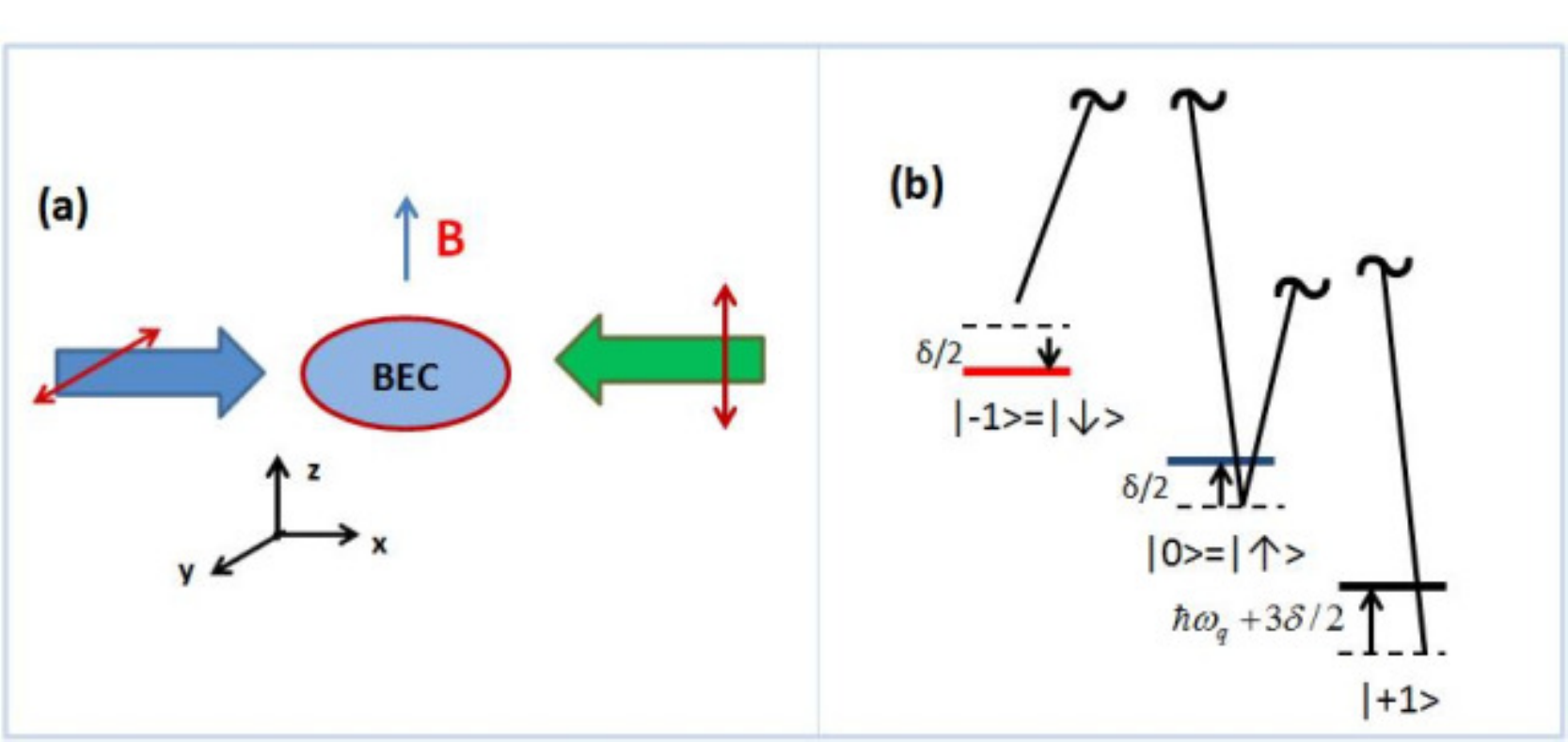} 
\caption{ (a) Schematic set up by NIST set up for exposing a BEC to two counter-propagating laser beam
(b) The Raman coupling between three hyperfine states.}
\label{raman}
\end{figure}

To generate SO coupling one considers the $^{87}$Rb atoms whose ground state electronic structure is $^{2}$S$_{1/2}$ giving  electron spin is $S=1/2$ and nuclear spin is $I=3/2$. Therefore the total spin $F$ 
can take value  $F=1$ and $F=2$ due to hyperfline coupling. The low energy manifold therefore consists of 
three $F=1$ states. Such states are characterized by by state vectors $|F,m_{F} \rangle$ which represents simultaneous eigenstates of $F^{2}$ and $F_{z}$ operators and for $F=1$, they are respectively given as $|1,1 \rangle$, $|1, 0 \rangle$ and $|1, -1 \rangle$. A schematic for the set-up is given in Fig. \ref{raman}. In presence of Zeeman field all these three levels that have same 
energy will split into three different levels. The resultant system is exposed to two counter propagating Raman laser beams along the $\hat{x}$ direction. The atom which is moving with velocity $\frac{\hbar k_{x}}{m}$ along 
$\hat{x}$ direction will absorb a photon coming from the oposite direction  of Laser I and will have momentum $\hbar( k_{x}-k_{L})$. From this excited state it will emit a a photon in the direction of laser II. As a result finally 
the momentum of the state along the $x$-direction will be $\hbar(k_{x} - 2k_{L})$ . 
Therefore the state will finally 
be written as $|-1, k_{x} - 2k_{L} \rangle$. Here the first quantum number 
corresponds to the hyperfine quantum number $m_{F}$, whereas the second one gives the momentum along the 
$\hat{x}$ direction.Similarly the atoms absorbing photon from laser II and emitting a photon in the direction of laser 
I will be finally in the state $|1, k_{x} + 2k_{L} \rangle$.
The final outcome is to have the following three states 
\bea |1 \rangle & = & | 1, k_{x} + 2k_{L} \rangle \nn \\
        |0 \rangle  & = & |0, k_{x} \rangle \nn \\
        |-1 \rangle & = &  |0, k_{x} -2k_{L} \rangle \eea 
It is possible to write down the effective hamiltonian now in $3 \times 3$ matrix form that includes contribution from 
atom, field ( laser) and atom-laser interaction.
However as has been shown explicitly in \cite{zhai}, it is possible to to tune the Zeeman energy and the laser frequency in such a way, that the low energy subspace created by $|1,0\rangle$ and $1,-1$ states 
are well separated from the $|1,1 \rangle$ state. Under this situation it is possible to construct an effective Hamiltonian in this two dimensional Hilbert space spanned by hyperfine state $|0 \rangle$ and $| -1 \rangle$ which we shall respectively call as  spin up $|\uparrow \rangle$ and spin-down $|\downarrow \rangle$ state. 
This is somewhat modified version of the scheme suggested in section \ref{NonAbel}. The $2 \times 2$ effective hamiltonian becomes, 

\beq H =  \begin{bmatrix}
    \frac{k_{x}^{2}}{2m}+\frac{\delta}{2} & \frac{\Omega}{2}e^{2ik_{L}x}\\
    \frac{\Omega}{2}e^{-2ik_{L}x} &  \frac{k_{x}^{2}}{2m}-\frac{\delta}{2}
  \end{bmatrix} \eeq 
Here $2 k_{0}$ is the momentum transfer due to the relative motion between the laser and the hyperfine state of 
the atom and $\delta$ is the detuning between the Raman resonance and the energy difference between the spin up and spin-down level. We have also absorbed an overall $\hbar$ factor in various terms. 
We refer to Ref. \cite{zhai} for the details about the derivation of the above hamiltonian. 

If one makes an unitary trasnformation on the two component wave function that will describe this system
such that  $\psi'=U\psi$, with
\beq  U=  \begin{bmatrix}
    e^{-ik_{L}x} & 0\\
    0 & e^{ik_{L}x}\\
  \end{bmatrix} \nn \eeq 

This changes the Hamiltonian from $H$ to $ UHU^{\dag}$ which is given by
\beq  H_{SO} =  \begin{bmatrix}
    \frac{(k_{x}+k_{L})^{2}}{2m}+\frac{\delta}{2} & \frac{\Omega}{2}\\
    \frac{\Omega}{2} &  \frac{(k_{x}-k_{L})^{2}}{2m}-\frac{\delta}{2}\\
  \end{bmatrix}  \label{SOH} \eeq 

\begin{figure}
\centering
\includegraphics[width=0.85\columnwidth , height= 
0.65\columnwidth]{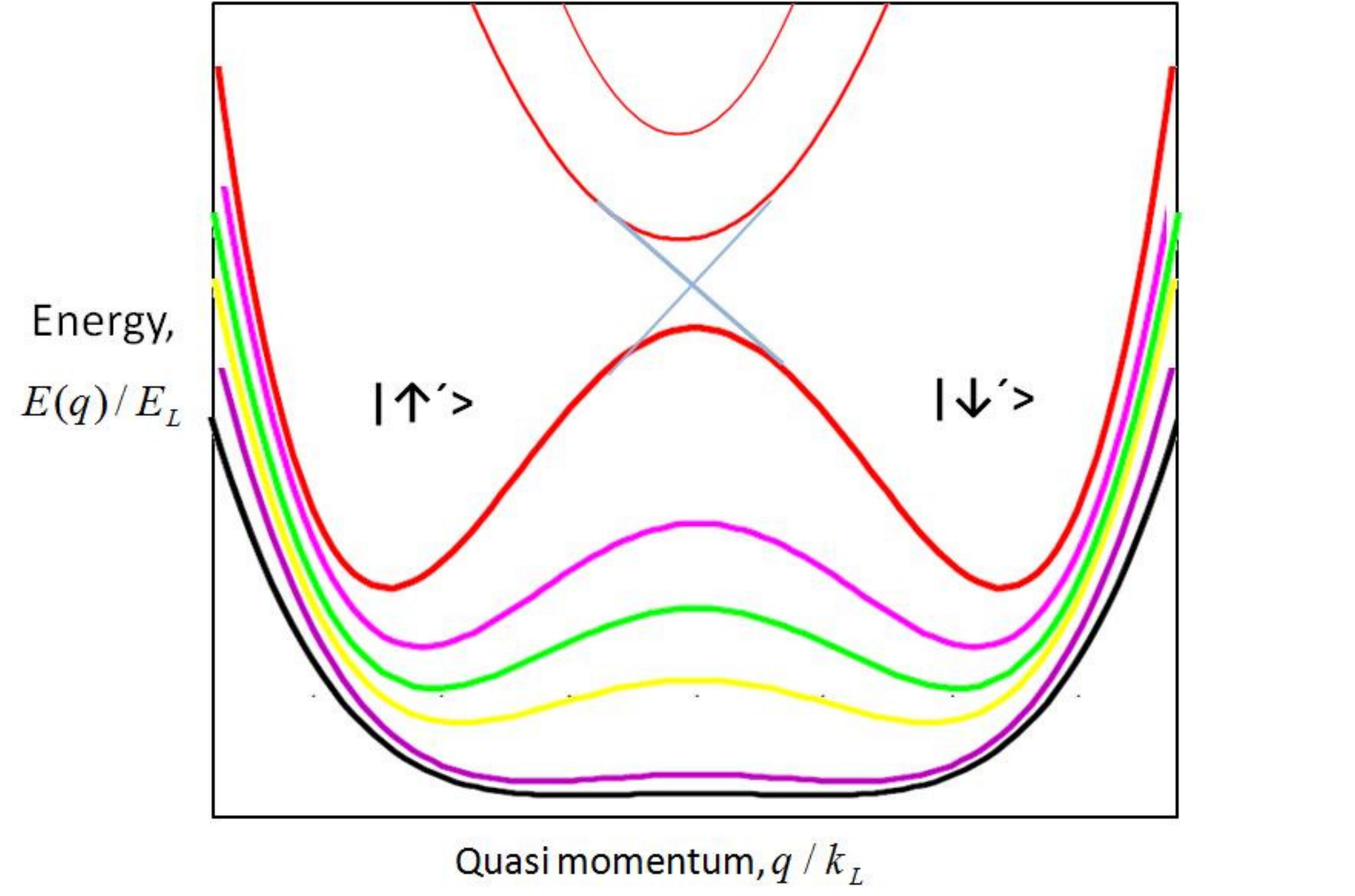} 
\caption{ The transition from a single to double non-denerate minima with the variation of detuning
for a spin orbit coupled gas. The idea of the Fig. is taken from \cite{lin1}.}
\label{energy}
\end{figure}

As one can see the resulting Hamiltonian can be written as 
\beq H_{SO}= \frac{(k_{x}\mathbb{I} + k_{L} \sigma_{z})^{2}}{2m} + \frac{\Omega}{2}\sigma_{x} + \frac{\delta}{2}\sigma_{z} \label{HSO1} \eeq 
The above hamiltonian is the spin-orbit coupled hamiltonian realized in NIST Experiment \cite{lin1}.
Even though here the vector potential has only one component $A_{x}$, however since that does not 
commute with the scalar potential $\frac{\Omega}{2}\sigma_{x} + \frac{\delta}{2} \sigma_{z}$, this is one of 
simplest realization of uniform non Abelian gauge potential. With a suitable spin rotation it can also be shown 
that the first term actually represent and linear combination of equal weight Rashba and Dresselhaus SO coupling. 
The energy eigenvalues are 
\beq \epsilon_{k} = \frac{k_{x}^{2} + k_{L}^{2}}{2m} \pm \sqrt{ (\frac{k_{x}k_{L}}{m} -\frac{\delta}{2})^{2}  + \frac{\Omega^{2}}{4} } \label{detuning}
\eeq 

which for  zero detuning $\delta=0$, the momentum dependent energy eigenvalues can be given as 
\beq \epsilon_{k} = \frac{k_{x}^{2} + k_{L}^{2}}{2m} \pm \sqrt{ (\frac{k_{x}k_{L}}{m})^{2}  + \frac{\Omega^{2}}{4} }
\eeq 
The condition for the minima of the energy can now be obtained from $\frac{\partial \epsilon_{k}}{\partial k_{x}}=0$
which gives one 
\bea  k_{x} & = & 0 \nn \\
 \text{or}~~~ k_{x} & = & \pm k_{0} \sqrt { 1 - (\frac{\Omega}{4E_{L}})^{2} }  \label{minima} \eea
Here $E_{L}= \frac{k_{L}^{2}}{2m}$
One can now checked the following things from the above expression. If the detuning $\delta$ is $0$, there will be 
distinctly one minima for $\Omega > 4E_{L}$ at $k_{x}=0$ and another minima at $k_{x}=\pm k_{0} \sqrt { 1 - (\frac{\Omega}{4E_{L}})^{2} } $ for $\Omega \le 4E_{L}$. However if the detuning $\delta$ is finite then whereas 
there is single minima at $k_{x}=0$ for $\Omega > 4E_{L}$, there are two non-degenrate minima at different height for 
$\Omega <  4E_{L}$. The height difference can be controlled by the detuning and the transition from single to 
two minima is the way one can detect the spin-orbit coupling \cite{lin1}. For Details of the experimental method 
one can look at \cite{Spielman, zhai}. A schematic of the variation of the energy $E$ as a function of the wavevector $q=k_{x}$ is given in Fig. \ref{energy}.

\section{Conclusion}
In this introductory review we provided a careful comparison between the true gauge fields that is responsible for the fundamental interaction between elementary particles and the syntehtic gauge field for ultra cold atoms. 
We analysed both Abelian and Non Abelian gauge field for this purpose. 
We have particularly show how both fundamental and synthetic gauge field can be interpreted in a similar geometric way; however the later does not have any independent dynamics and hence dubbed as synthetic.
We also illustrate the examples of such synthetic gauge field for ultra cold atoms by considering two specific cases:  Abelian gauge field in rotating Bose Einstein condensates and Non Abelian gauge field in spin-orbit coupled Bose Einsten condensates. This primer by no way covered the large amount of exciting work that was done in the field of synthetic gauge field for ultra cold atoms. We referred to a number of excellent review articles to that purpose. 
We also cited only a limited number of mostly pedagogical articles and books on the relevant topic and apologize for our inability to cite a large number of exciting and important and highly relevant work in this field. 
We hope  the direction and information given in this review will be sufficient to direct the interested reader to more 
complete set of references on synethetic gauge field. 

One of us (SG) had an opportunity to give a set of letures on "Cold Atoms in Artificial Gauge field" in a Winter School in IISER Pune in December, 2013 alongsider with a set of lectures given on "Abelian and Non Abelian Gauge Fields" by Prof. Sunil Mukhi.  SG takes this opportunity to thank Prof. Sunil Mukhi for explaining 
the Abelian and Non Abelian gauge theory in an extremely lucid way which benefitted him a lot.

\end{document}